\documentclass[prl,amsmath,aps,floats,amssymb, floatfix, superscriptaddress, twocolumn,nofootinbib]{revtex4}
\usepackage{graphicx}
\usepackage{wasysym}
\usepackage{amsfonts}
\usepackage{subfigure}
\usepackage{hyperref}
\usepackage{color}        
\usepackage{ulem}
\usepackage{lipsum}

\newcommand{\sfig}[2]{
\includegraphics[width=#2]{#1}
        }
\newcommand{\Sfig}[2]{
    \begin{figure}[thbp]
    \sfig{#1.pdf}{1.\columnwidth}
    \caption{{\small #2}}
    \label{fig:#1}
    \end{figure}
}

\newcommand{\rf}[1]{\ref{fig:#1}}

\newcommand\be{\begin{equation}}
\newcommand\ee{\end{equation}}

\def\eea{\end{eqnarray}}

\def\bea{\begin{eqnarray}}
\def\eea{\end{eqnarray}}

\begin{document}

\title{How much can we learn about the physics of inflation?}         
\author{Scott Dodelson}        
\affiliation{Fermi National Accelerator Laboratory, Batavia, IL 60510-0500}
\affiliation{Kavli Institute for Cosmological Physics, Enrico Fermi Institute, University of Chicago, Chicago, IL 60637}
\affiliation{Department of Astronomy \& Astrophysics, University of Chicago, Chicago, IL 60637}

\begin{abstract}
\noindent
The recent BICEP2 measurement of B-modes in the polarization of the cosmic microwave background suggests that inflation was driven by a field at an energy scale of $2\times 10^{16}$ GeV. I explore the potential of upcoming CMB polarization experiments to further constrain the physics underlying inflation. If the signal is confirmed, then two sets of experiments covering larger area will shed light on inflation. Low resolution measurements can pin down the tensor to scalar ratio at the percent level, thereby distinguishing models from one another. A high angular resolution experiment will be necessary to measure the tilt of the tensor spectrum, testing the consistency relation that relates the tilt to the amplitude.
 \end{abstract}

\newcommand\clb{C_l^{BB}}
\maketitle

\section{Introduction}

The BICEP2 experiment~\cite{Ade:2014gua,Ade:2014xna} has detected the B-modes of polarization in the cosmic microwave background (CMB), a signature~\cite{Seljak:1996gy,Kamionkowski:1996zd} of primordial gravitational waves generated during inflation~\cite{Guth,Starobinsky:1979ty,Linde,Mukhanov:1981xt,Albrecht}. If these results are confirmed and are indeed due to inflation, then the energy scale responsible for the early epoch of acceleration is $2\times 10^{16}$ GeV, some 13 orders of magnitude above energy scales probed in the largest colliders today. 

The result is stunning with far-ranging implications. Here I focus on what comes next: what can we learn about this new physics from the next generation or two of CMB polarization experiments? Although this topic has been addressed before~\cite{Song:2003ca,Verde:2005ff,Baumann:2008aq,Easson:2010uw,Abazajian:2013vfg,Wu:2014hta}, we are now (probably) in a new era, in which we know the amplitude of the signal so are standing on firmer ground. 

My conclusions are that we are about to embark on a three-step journey that can potentially uncover the laws of physics at ultra-high energies using CMB polarization:
\begin{itemize}
\item The first step is to confirm the BICEP2 result and determine the amplitude of the gravitational wave signal. There are several ways to do this. The most comforting would be a detection by the Planck satellite of the first B-mode peak on large scales. BICEP2 is not sensitive to this reionization-induced signal, so a detection would complement and confirm with little doubt that we have indeed observed primordial gravitational waves. Another test would be a detection with one of the other ground-based experiments, especially if the measurement were made in a different part of the sky and at a different frequency. In short, there are three axes along which we can move to confirm the BICEP2 result: angular frequency to detect the distinctive two-humped signal, photon frequency to eliminate the possibility of foreground contamination, and sky coverage again to mitigate foregrounds. For the next two steps to move forward, this first stage needs to conclude with the removal of the tensions between different data sets that predate not only BICEP2, but even Planck~\cite{Hou:2012xq,Archidiacono:2013lva}. If these tensions remain, the resolution may require ways of understanding the responsible physics beyond those outlined below; e.g. bumps in the primordial power spectrum~\cite{Contaldi:2014zua,Miranda:2014wga}. I will have nothing to say here about this stage, except the obvious: it is very important.

\item The second stage will be to measure the tensor to scalar ratio, $r$, and the spectral index of the scalar perturbations, $n_s$, with increasing accuracy. Models make predictions in the $n_s,r$ plane~\cite{Linde:1983gd,Freese:1990rb,Dodelson:1997hr,Silverstein:2008sg,McAllister:2008hb,Baumann:2008aq,Kaloper:2008fb,Kaloper:2011jz,Ade:2013uln}, and increasing precision can help identify the correct model. Indeed, a host of models are already ruled out if $r$ is close to 0.2 as suggested by the BICEP2 results. There are plans to reduce the errors on $n_s$ by a factor of five using galaxy surveys~\cite{Levi:2013gra}, and future CMB polarization experiments can reduce the error on $r$ to the percent level. As we will see (e.g., Fig.~\rf{contour_nc_r}), this can be done mostly by increasing the sky coverage (BICEP2 covers less than a percent of the sky) even with a relatively large beam. This is easier said than done of course because BICEP2 looked at a low-foreground region, so this next generation of experiments will likely need to be equipped with multiple frequency channels in order to disentangle the signal from the foregrounds. A very important physical question underlying model choice is why the ``simplest model,'' with the field driving inflation subject to a quadratic potential, seems to fit the data. Everything we know about effective field theory tells us that since the field traverses a large distance in Planck units, higher order terms should be generated in the effective action, completely changing the simple dynamics of an $m^2\phi^2$ term. Different solutions to this problem make different predictions in the $n_s,r$ plane,. Therefore, this ``{\it Simple fits but Simple doesn't make sense}'' quandary may be resolved by obtaining greater precision in the ($n_s,r$) plane.

\item The third stage will be to measure the running of the spectral index of the tensor perturbations, $n_t$, and test the prediction that $n_t=-r/8$~\cite{Caligiuri:2014sla}. As we will see (e.g, Fig.~\rf{contour_c5_nt}), carrying out this program will require exquisite ``cleaning'' of the B-mode signal from lensing~\cite{Hu:2001kj,Seljak:2003pn,Smith:2010gu} and therefore will require small-scale resolution, low-noise, and large sky coverage. Even if all these are achieved, the conclusion that $n_t$ will be measured to be non-zero rests on the assumption that $r$ is large. If we find in Step 1 that $r=0.1$, it will become virtually impossible (Fig.~\rf{best_five2r}) to detect non-zero $n_t$ at even the 2-sigma level, although there is still an enormous amount of physics that can still be gleaned from high resolution polarization experiments.
\end{itemize}

It is unlikely that these steps will occur sequentially: we should expect progress on all three fronts over the coming decade.

To quantify these conclusions, I project constraints from polarization experiments in the two dimensional $r,n_t$ plane. The errors on $C_l^{BB}$, sample variance and noise, are computed using the standard formula~\cite{Knox:1995dq,Wu:2014hta}. On small scales, this simple formula captures the noise reported by BICEP2~\cite{Ade:2014gua,Ade:2014xna} with sky coverage set to 384 square degrees; noise per square degree pixel set to $0.087\times \sqrt{2}\mu$K; and beam width equal to $30'$ FWHM. The formula underestimates the noise in BICEP2 at low $l$ probably because it does not account for low $l$ removal from filtering\footnote{Thanks to Chris Sheehy for suggesting this.}, so the estimates presented here may be a bit optimistic. On the other hand, the projected error on $r$ using this Fisher formalism is only 20\% smaller than that obtained by BICEP2 in their analysis. As we will see, a twenty percent difference does not matter much for the calculations in the next section, 
where the marginalized constraints on $r$ are displayed, and in following section, 
where the constraints on $n_t$ are examined, most of the weight comes from small scales where the formula agrees well with the BICEP2 errors.

\newcommand\fsky{f_{\rm sky}}

\section{Inflationary models}
\label{sec:infl}

By way of orientation, let us first consider the constraints on the tensor to scalar ratio $r$ (after marginalizing over the tilt $n_t$) from a perfect CMB polarization experiment that covers the whole sky with zero instrumental noise. Fig.~\rf{bestr_nc2r} projects the constraints on $r$ from such an experiment as a function of the maximum $l$ used (the smallest scale) and the minimum $l$. The monotonically increasing solid blue curve shows that the constraint on $r$ comes from multipoles $l<150$. This flattening is due to noise from the lensed E-modes, which is not cleaned in Fig.~\rf{bestr_nc2r}. The take-away is that -- even with no cleaning -- an all-sky, low noise, low resolution experiment could obtain percent level constraints on $r$. The dashed curve shows that this conclusion changes quantitatively but not qualitatively if $r$ is smaller than the central value of 0.2 reported by BICEP2. The information at low $l$ is cosmic variance limited so $r/\Delta r$ is independent of $r$ in that regime. The red monotonically decreasing curves in Fig.~\rf{bestr_nc2r} shows that percent level accuracy is still possible even if the first hump at $l<10$ is not measured. That is, even if $l_{\rm min}$ is of order 50, the constraints on $r$ would still be tighter than 2\%, again in this ideal case. 

\Sfig{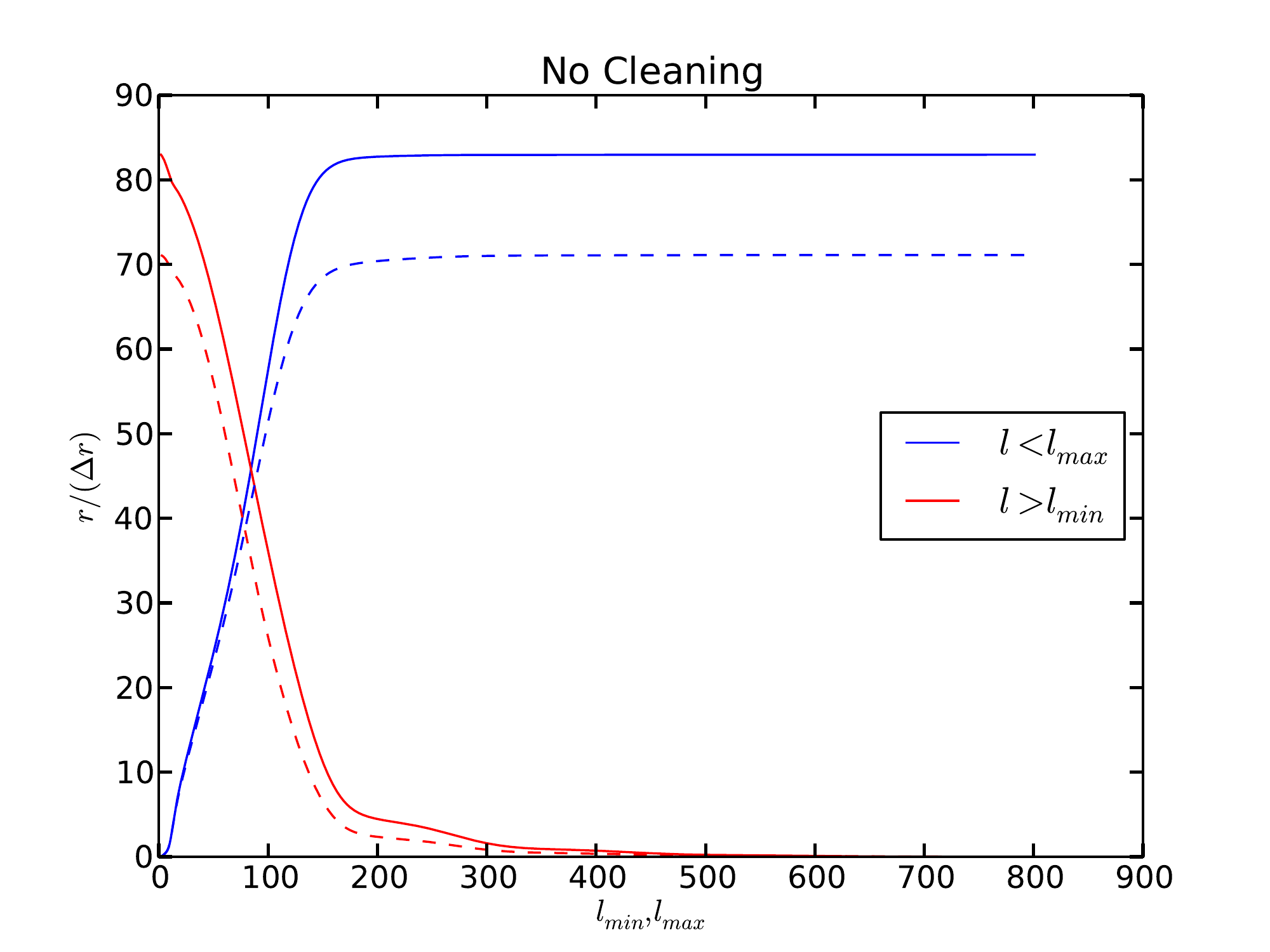}{Optimal limit on $r$ assuming full sky coverage and zero instrumental noise if there is no cleaning of the lensing contamination. Dashed curves are for $r=0.1$; solid for $r=0.2$. Monotonically increasing blue curves show the constraints if only multipoles $l<l_{\rm max}$ are used; apparently multipoles great than 150 do not contribute to the constraints. Monotonically decreasing red curves show the constraint if only $l>l_{\rm min}$ are used, showing that the large scale reionization bump at $l<10$ is not crucial for obtaining tight constraints on $r$.}

There are a variety of ways to clean the noise created by the lensed E-modes. Most powerful is to measure the polarization on small scales, estimate the projected gravitational signal from these measurements, and then subtract off the synthetic lensed E-modes. There are other ways to estimate the gravitational potential, for example from the CMB temperature field~\cite{Das:2011ak,vanEngelen:2012va,Ade:2013tyw} or from galaxy surveys~\cite{Hanson:2013daa}. A number of groups~\cite{Hu:2001kj,Seljak:2003pn,Smith:2010gu} have argued that cleaning the lensed E-modes with internal small scale polarization maps or external maps of large scale structure could reduce the lensing noise (the amplitude squared) by a factor of ten or better.  Fig.~\rf{bestr_five} shows the ensuing projections, again for an all-sky-- no noise hypothetical experiment, if the lensed E-mode spectrum could be reduced by a factor of twenty. The signal to noise would go up a bit but the general conclusion that only large scales contribute to the constraints remains unchanged.

\Sfig{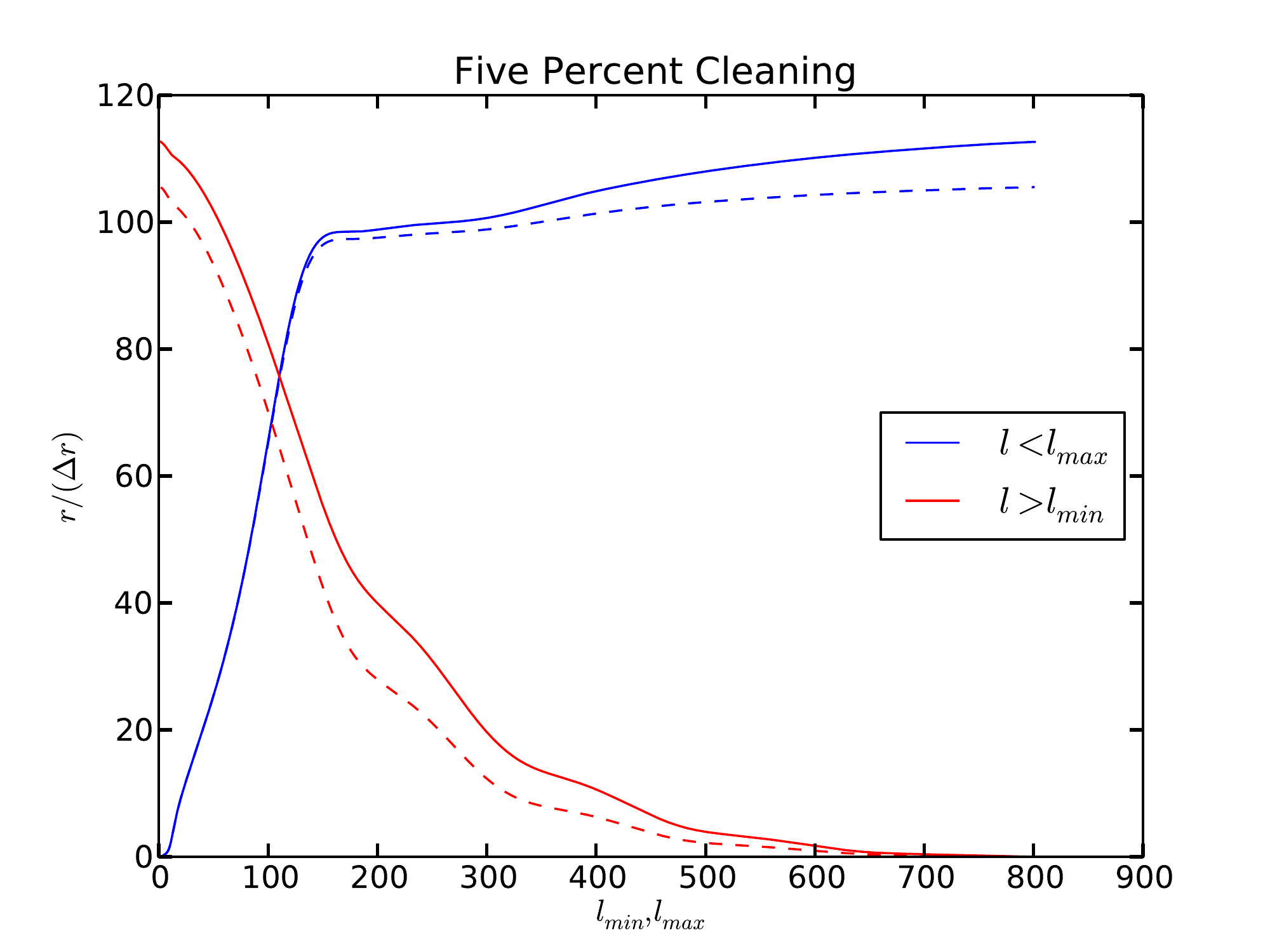}{Optimal limit on $r$ assuming full sky coverage and zero instrumental noise if there is 5\% cleaning of the lensing contamination.}

We are currently far from this ideal experiment. Fig.~\rf{contour_nc_r} shows what is needed to get to the percent level. Most important is to obtain more sky coverage. The BICEP sensitivity is adequate as long as more of the sky is measured. If foregrounds were not a problem, then a hundred copies of the BICEP2 configuration would suffice to obtain a measurement of $r$ with two percent errors. Finding the optimal frequency coverage and experimental configuration to measure more of the sky will likely consume the community as the dust settles on the BICEP2 result.

\Sfig{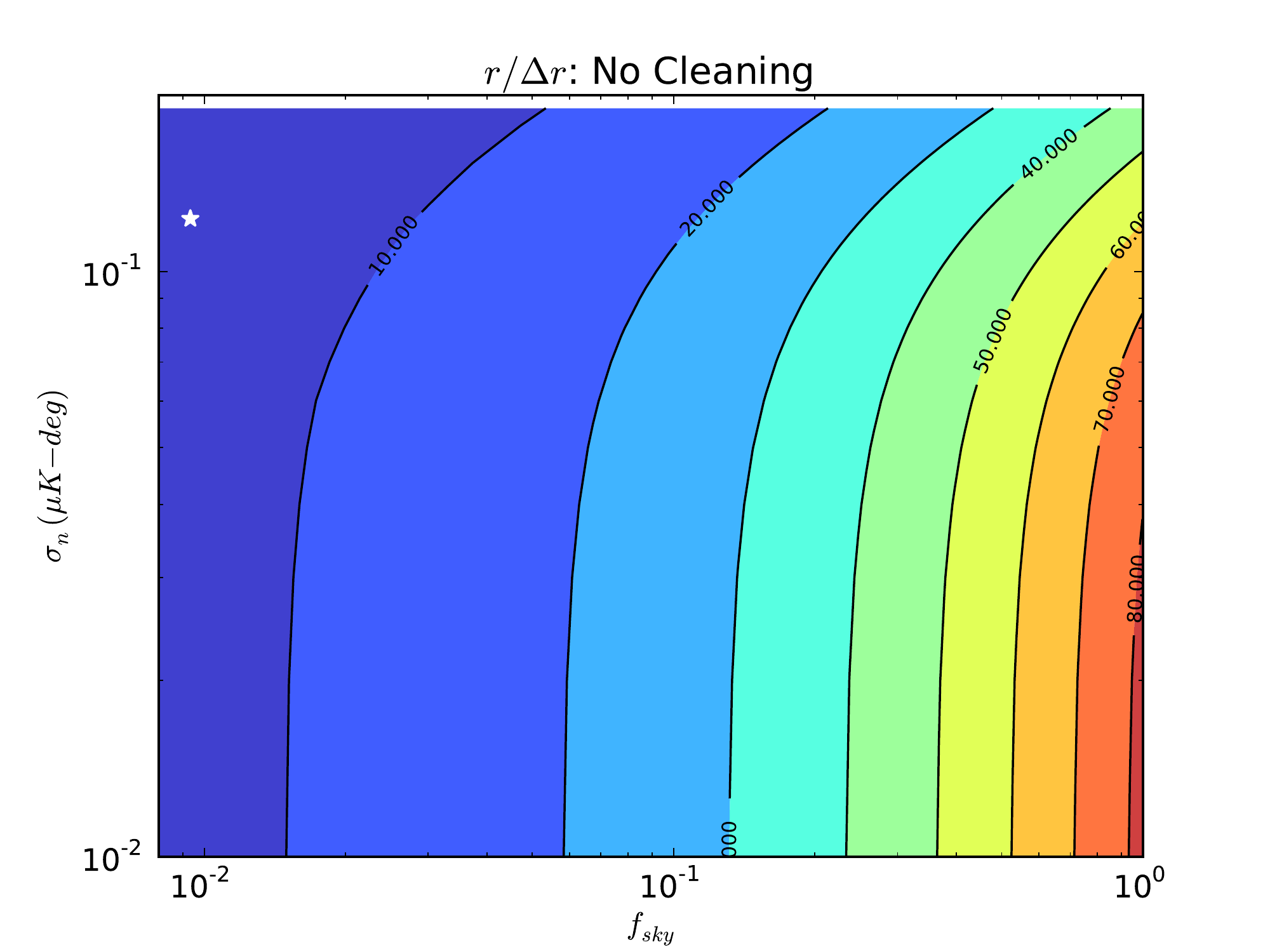}{Signal to noise on $r$ as a function of pixel noise and sky coverage. The BICEP2 specs are in the upper left at the starred point. Here $r=0.2$ is assumed.}

Fig.~\rf{contour_beam2bic_r} shows that the beam is one parameter that is not essential. A degree beam (twice the size of the BICEP2 beam) would produce constraints comparable to those shown in Fig.~\rf{contour_nc_r}, where the beam was assumed to have infinite resolution. This follows directly from the observation that most of the information on $r$ lies in the multipole range $l<150$. Apart from cleaning the lensing signal, there is no reason to measure on very small scales if the goal is to obtain tight constraints on $r$.

\Sfig{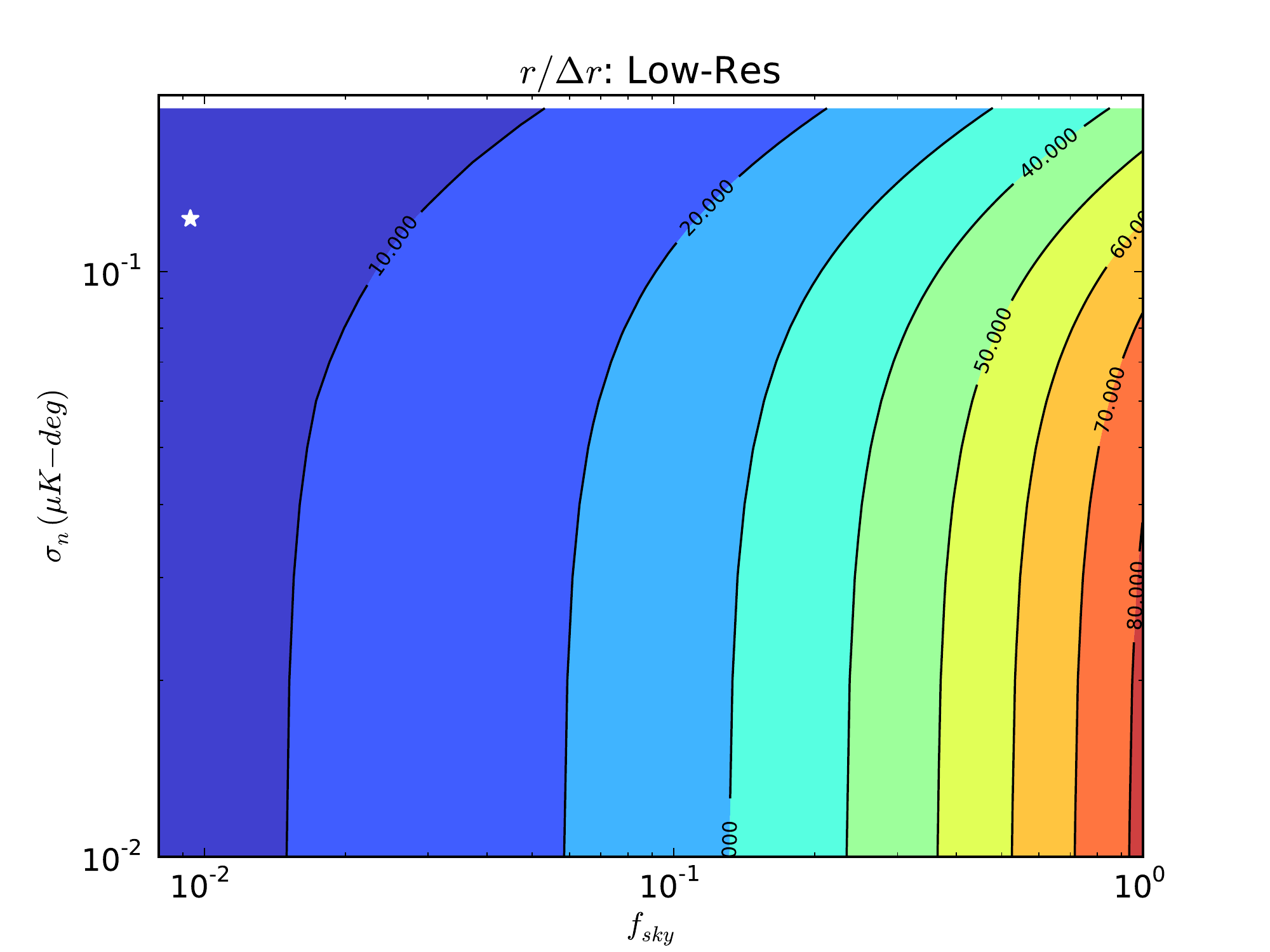}{Same as Fig.~\rf{contour_nc_r} but with a low-resolution beam $\theta_{\rm FWHM}$ twice as large as that employed in the BICEP2 experiment.}

\section{Running}\label{nt}

Measuring the amplitude of the gravitational wave signal is qualitatively different than measuring its spectral shape. We quantify the shape of the spectrum with a power law index, so that $k^3 P_{\rm gw}(k)\propto k^{n_t}$. In slow roll models of inflation that are driven by a single scalar field, both $r$ and $n_t$ are proportional to $(V'/V)^2$ where $V$ is the potential of the inflation field and $V'$ is its derivative with respect to the field. So in most models, $n_t$ is predicted to deviate from zero. Depending on the scale at which $r$ is evaluated\footnote{Here, the tensor to scalar ratio is evaluated at $k=0.009$ h Mpc$^{-1}$ since that is the scale that determines the height of the peak at $l=80$.}, the proportionality constant is fixed; $n_t=-r/8$. 

\Sfig{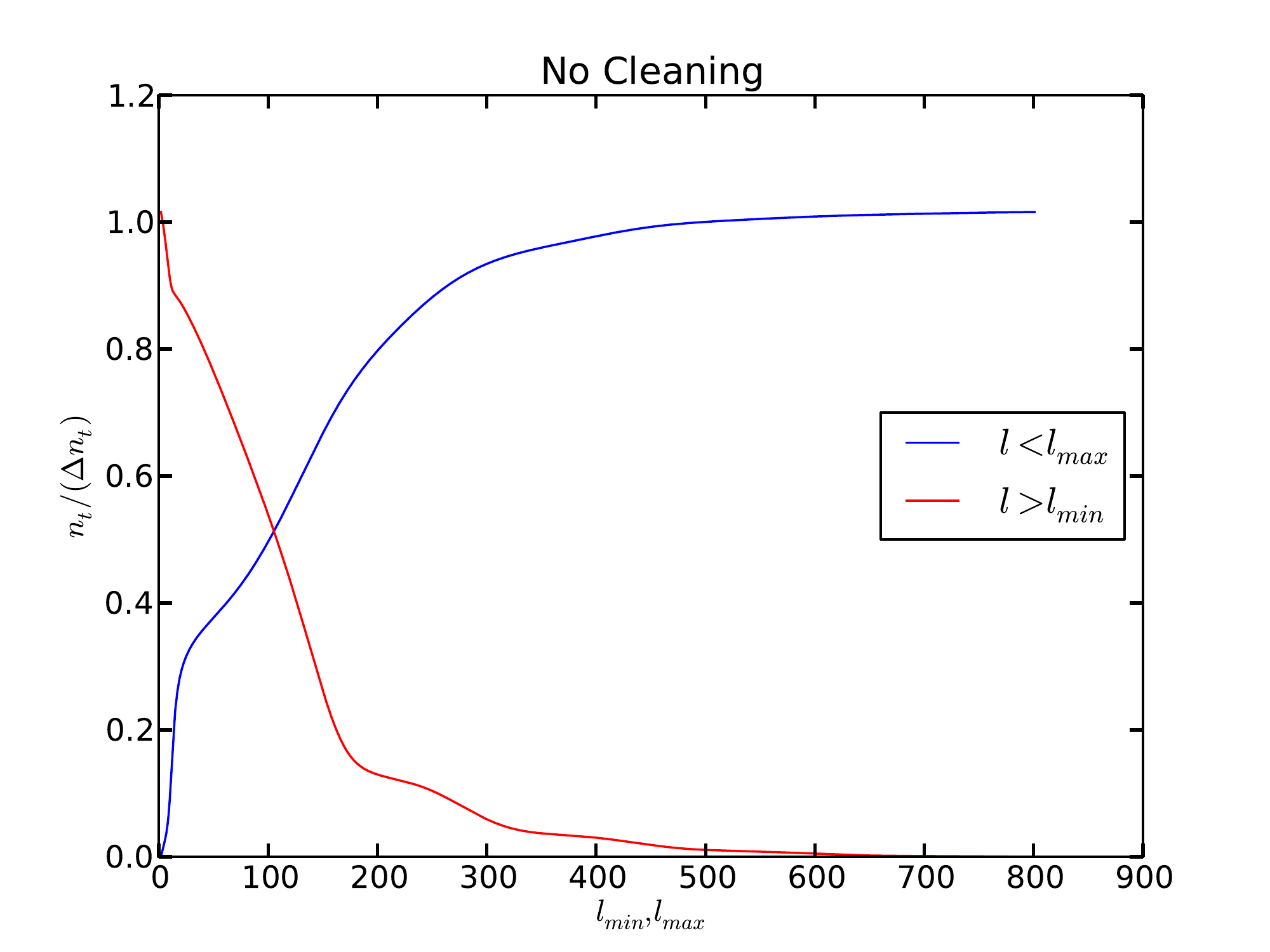}{Optimal limit on $\Delta n_t$ assuming full sky coverage and zero instrumental noise if there is no cleaning of the lensing contamination.}

Measuring the shape of the spectrum requires a long lever arm so a wider range of angular scales is necessary. This can be seen in Fig.~\rf{best_five2r} where -- in contrast to information localized at $l<150$ for the amplitude -- mulitpoles as large as $l\sim300$ contribute to the signal to noise on $n_t$. Even for this optimal -- all-sky, no noise -- idealization, the detection of non-zero $n_t$ would be at only 1-sigma. The contamination from the lensed E-modes is particularly damaging as it impedes the use of the smaller angular scales necessary to measure the slope of the spectrum.

\Sfig{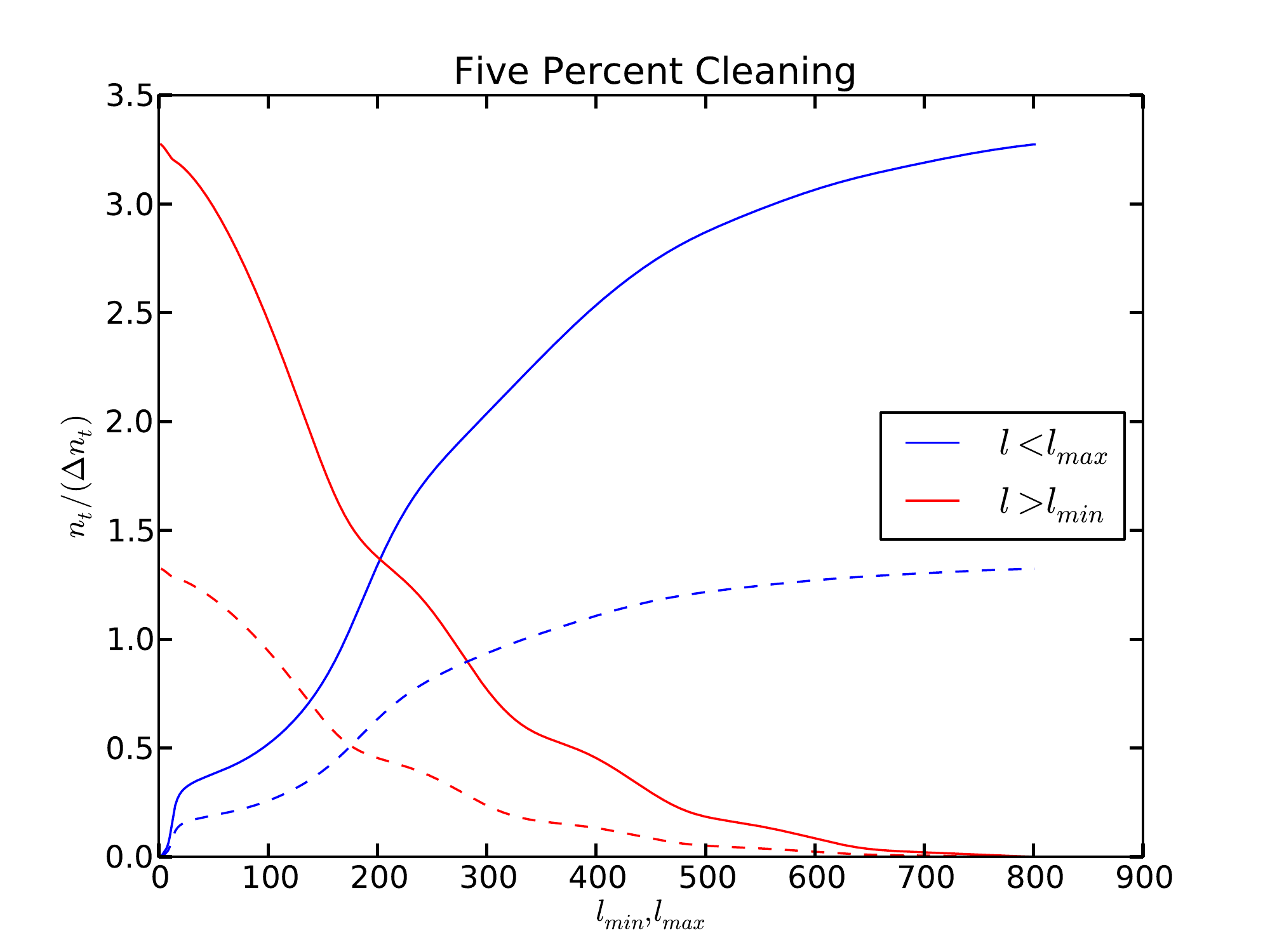}{Same as Fig.~\rf{best_nc}, but this time assuming the lensed signal can be cleaned at the 5\% level. Dashed curves show the projected constraint on $n_t$ if the true value of r is $0.1$.}


Fig.~\rf{best_five2r} shows how the situation improves if the lensing contamination can be removed. If 95\% of the lensing spectrum can be removed, then Fig.~\rf{best_five2r} shows that the projected error on $n_t$ moves above 3-sigma, again for this optimal configuration. 
An important feature of this plot is that a significant detection depends sensitively on the value of $r$. The dashed curves show that, even with five percent cleaning in an all-sky configuration, CMB polarization cannot detect at more than 1-sigma the deviation of $n_t$ from zero if the true value of $r$ is 0.1.

\Sfig{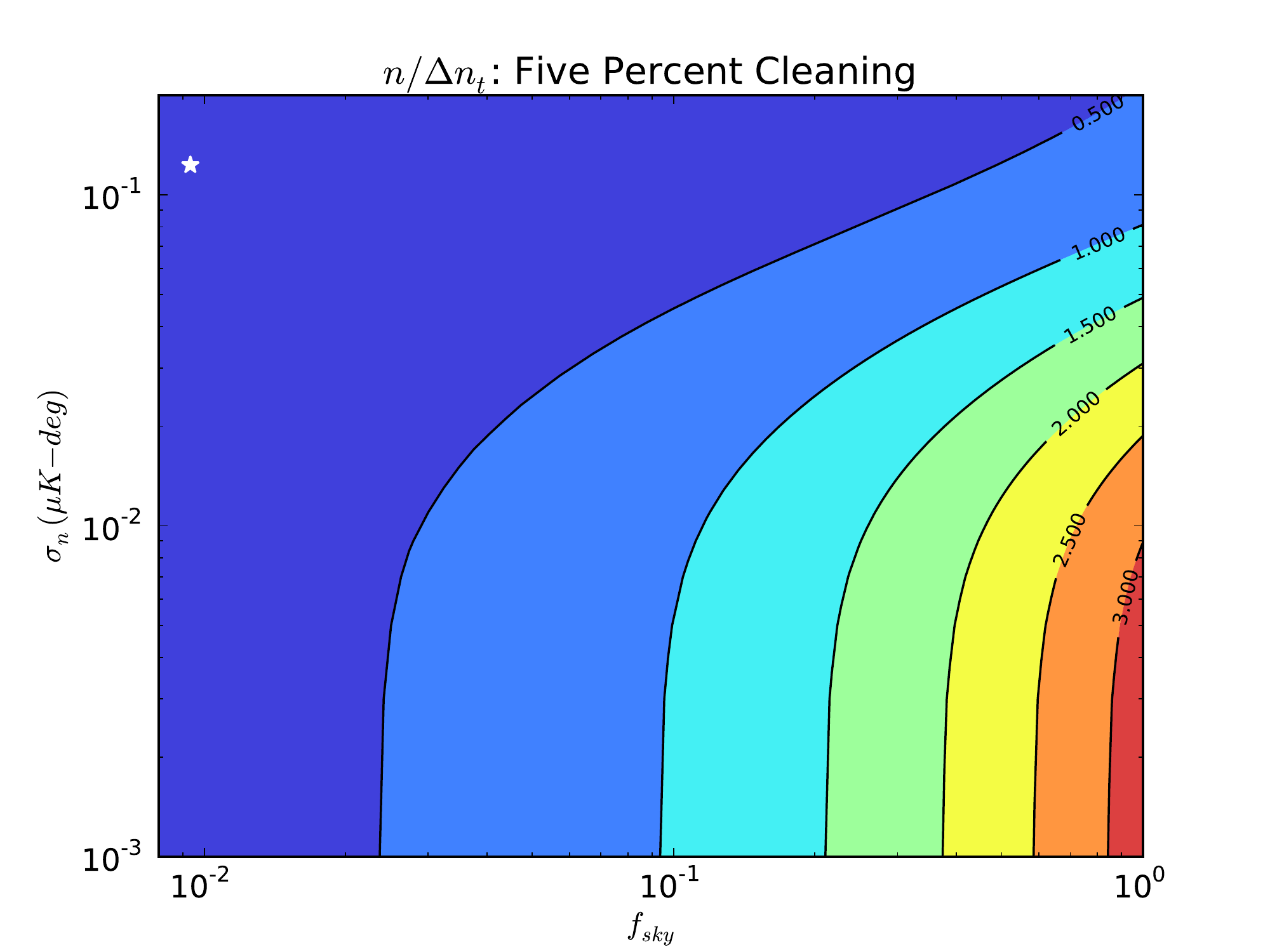}{Detectability of the tensor slope: $\Delta n_t/n_t$ as a function of pixel noise and sky coverage. Plot assumes $r=0.2$ and five percent cleaning of the lensing spectrum. The star denotes the BICEP2 configuration.}

Fig.~\rf{contour_c5_nt} shows how far we have to go before reaching this signal to noise. A fair fraction of the sky must be covered with exquisite sensitivity, an order of magnitude more sensitivity than BICEP2.


{\it Acknowledgments} I am grateful to Tom Crawford, Matthew Dodelson, Eiichiro Komatsu, Marilena Loverde, Chris Sheehy, and Anze Slosar for useful suggestions and conversations. This work is supported by the U.S. Department of Energy, including grant DE-FG02-95ER40896. {\bf Note added}: after this work was finished, the preprint \cite{Caligiuri:2014sla} appeared, which has overlap with these results.

\bibliography{inflation}

\end{document}